\documentclass[aps,prep,preprintnumbers,epsf,amsmath,amssymb,nofootinbib]{revtex4}
\usepackage{amssymb}
\usepackage{amsmath}
\usepackage{bm}
\usepackage{graphicx}
\usepackage{epsfig}
\draft
\begin{document}
\title{The primordial explosion of a false white hole from a 5D vacuum}
\author{$^{1}$ Jos\'e Edgar Madriz Aguilar \footnote{E-mail address:edgar.madriz@red.cucei.udg.mx},
$^{1}$ Claudia Moreno \footnote{E-mail address:
claudia.moreno@cucei.udg.mx}
 and $^{2,3}$ Mauricio Bellini
\footnote{E-mail address: mbellini@mdp.edu.ar}, }
\affiliation{$^{1}$ Departamento de Matem\'aticas, Centro
Universitario de Ciencias Exactas e ingenier\'{i}as (CUCEI),
Universidad de Guadalajara (UdG), Av. Revoluci\'on 1500 S.R. 44430, Guadalajara, Jalisco, M\'exico.  \\
and \\
$^{2}$ Departamento de F\'isica, Facultad de Ciencias Exactas y
Naturales, Universidad Nacional de Mar del Plata (UNMdP),
Funes 3350, C.P. 7600, Mar del Plata, Argentina \\
E-mail: mbellini@mdp.edu.ar\\
$^{3}$ Instituto de Investigaciones F\'{i}sicas de Mar del Plata
(IFIMAR)- Consejo Nacional de Investigaciones Cient\'{i}ficas y
T\'ecnicas (CONICET) Argentina.}

\begin{abstract}
We explore the cosmological consequences of some possible big bang
produced by a black-hole with mass $M$ in an 5D extended SdS.
Under these particular circumstances, the effective 4D metric
obtained by the use of a constant foliation on the extra
coordinate is comported as a false white-hole (FWH), which
evaporates for all unstable modes that have wavelengths bigger
than the size of the FWH. Outside the white hole the repulsive
gravitational field can be considered as weak, so that the
dynamics for fluctuations of the inflaton field and the scalar
perturbations of the metric can be linearized.
\end{abstract}

\maketitle

\section{Introduction and Motivation}

In the last two decades, inflationary cosmology has become the
strongest candidate to explain the observed properties of the
universe on cosmological scales\cite{0,1,2,3,4,5}. This fact is
supported by experimental evidence\cite{smoot}. During this epoch
the energy density of the Universe was dominated by some scalar
field (the inflaton), with negligible kinetic energy density, in
such a way that its corresponding vacuum energy density is
responsible for the exponential growth of the scale factor of the
universe. Along this second order phase transition a small and
smooth region of the order of size of the Hubble radius, grew so
large that it easily encompassed the comoving volume of the entire
presently observed universe, and consequently the observable
universe become so spatially homogeneous and isotropic on scales
today of the range [$10^8-10^{10}$] ly. There are plenty of
inflationary models\cite{encicl} the majority of them in good
concordance with observations, but none free of problems, as for
example the transplanckian problem, the hierarchy problem, etc
\cite{bran}. This has lead cosmologists to look for some new
theoretical alternatives \cite{nrr}.

Recently, we have suggested a new inflationary
perspective\cite{cos} from an 5D Extended Theory of General
Relativity (ETGR)\cite{etgr} in which we use the Ricci-flat
metric\cite{etgr} represented by a 5D line element\footnote{Some
related metrics were studied in \cite{etgr1}.}
\begin{equation}\label{a1}
dS_{5}^{2}=\left(\frac{\psi}{\psi_0}\right)^{2}\left[c^{2}f(R)\,dT^2
-\frac{dR^2}{f(R)}-R^2(d\theta^2+sin^2\theta\,
d\phi^2)\right]-d\psi^2,
\end{equation}
Here, $f(R)=1-[(2G\zeta\psi_0/(Rc^2)]-(R/\psi_0)^2$ is a
dimensionless metric function, $\psi$ is the space-like and
non-compact fifth extra coordinate\footnote{In our notation
conventions henceforth, latin indices $a,b=$ run from 0 to 4,
whereas the rest of latin indices $i,j,n,l,...=$ run from 1 to
3.}. The metric (\ref{a1}) is Ricci flat. Furthermore, this is an
extension to 5D spaces of the 4D SdS metric. The coordinate $T$ is
a time-like, $c$ is denoting the speed of light, $R,\theta,\phi$
are the usual spherical polar coordinates, $\psi_0$ is an
arbitrary constant with length units and the constant parameter
$\zeta$ has units of $(mass)(length)^{-1}$. As was shown in
\cite{etgr}, for certain values of $\zeta$ and $\psi_0$, the
metric in (\ref{a1}) has two natural horizons. The inner one is
the analogous to the Schwarzschild horizon and the external one is
the analogous to the Hubble horizon. A particular case of the
metric (\ref{a1}) is such that $\zeta=1/(3\sqrt{3} G)$. For this
case there is an unique horizon at $R_*=\psi_0/\sqrt{3}$, which is
false, because $f(R)\leq 0$ retains its sign for $R\lessgtr R_*$.
However the metric (\ref{a1}) really fails at $R=0$. It can be
seen more clear by writing $f(R)$ in this special case as
\begin{equation}
f(R)= \left(-\frac{1}{\psi^2_0 R}\right)
\left[\left(R+\frac{2\psi_0}{\sqrt{3}}\right)\left(R -
\frac{\psi_0}{\sqrt{3}}\right)^2\right].
\end{equation}
Of course the physical domain of interest is $R>0$. It is easy to
see that $F(R_*)=0$, but however $dS_{5}^{2}\leq 0$ has the same
signature for $R\lessgtr R_*$. As was demonstrated in \cite{etgr},
the Newtonian induced acceleration in absence of angular moment
reads
\begin{equation}\label{in2}
a_c=-\frac{\psi_0}{3 \sqrt{3} R^2}+\frac{Rc^2}{\psi_{0}^2}.
\end{equation}
This means that, for the special case $\zeta=1/(3\sqrt{3} G)$, the
acceleration (\ref{in2}) becomes zero at $R=R_*$. Moreover, this
acceleration is negative for $0<R<R_*$ and positive for $R>R_*$.
In other words, the metric (\ref{a1}) predicts that for
$\zeta=1/(3\sqrt{3} G)$, gravitation is repulsive for $R> R_*$ and
attractive for $R<R_*$. As we shall see, this has important
cosmological consequences. In physical terms, the case we are
dealing to study in this work is very important because describes
the greatest possible mass which can have a black-hole,
$M=1/(3\sqrt{3} G) \psi_0$, in an universe with cosmological
constant $3/\psi^2_0$. The metric (\ref{a1}) is static, however,
it can be written on a dynamical coordinate chart $\lbrace
t,r,\theta,\phi,\psi\rbrace$ by implementing the planar coordinate
transformation \cite{pc}. For the case $M=1/(3\sqrt{3} G) \psi_0$,
the transformations are
\begin{equation}\label{a2}
R=ar\left[1+\frac{\psi_0}{6\sqrt{3} ar}\right]^2,\quad T=t+H\int^r
dR\frac{R}{f(R)}\left(1-\frac{2\psi_0}{3\sqrt{3} R}\right)^{-1/2},
\end{equation}
$a(t)=a_0e^{t/\psi_0}$ being the scale factor. With this
transformation the line element (\ref{a1}) reads
\begin{equation}\label{a3}
  dS_{5}^{2} = \left(\frac{\psi}{\psi_0}\right)^2 \left[F(\tau,r)d\tau^2 - J(\tau,r)(dr^2+r^2(d\theta^2+sin^2\theta
  d\phi^2))\right]-d\psi^2,
\end{equation}
where $\tau$ is the conformal time, $d\tau=a^{-1}(\tau)dt$ and
$a(\tau)=-\psi_0/\tau$. Furthermore, the metric functions
$F(\tau,r)$ and $J(\tau,r)$ are given by
\begin{equation}\label{a4}
F(\tau,r)=a^2(\tau)\left[1-\frac{\psi_0}{6\sqrt{3}
a(\tau)r}\right]^2\left[1+\frac{\psi_0}{6\sqrt{3}
a(\tau)r}\right]^{-2},\quad
J(\tau,r)=a^2(\tau)\left[1+\frac{\psi_0}{6\sqrt{3}
a(\tau)r}\right]^4.
\end{equation}
The constant Hubble parameter satisfies
\begin{equation}\label{a5}
H=\frac{1}{\psi_0}=a^{-2}\frac{da}{d\tau} .
\end{equation}
The acceleration $a_c$, can be written in terms of the new
coordinates $r$ and $\tau$
\begin{equation}\label{ac}
a_c(r,\tau) = \frac{324 \sqrt{3} r^2 \tau - 1944 r^3 - 215
\sqrt{3} \tau^3 - 54 r \tau^2}{6 \psi_0 \tau (\sqrt{3} \tau - 18 r
)^2}.
\end{equation}
The physical distance for which the acceleration (\ref{ac})
becomes null, which provide us the size of the FWH in a comoving
frame, is
\begin{equation}
D_{wh} = a(\tau_{wh}) \,r_{wh} = \frac{5 \sqrt{3}}{18} \psi_0,
\end{equation}
so that $a_c(D_{wh})=0$. The  wavenumber corresponding to the
scales $D_{wh}$ is
\begin{equation}\label{kga}
k_{wh}=\left[\frac{2\pi}{D_{wh}}\right]\left[\frac{2D_{wh}}{2D_{wh}
+\frac{\psi_0}{3\sqrt{3}}}\right]^{2},
\end{equation}
give us the wavelength related to $k_{wh}$
\begin{equation}\label{b6}
\lambda_{wh} = \frac{2\pi}{k_{wh}} \simeq 0.6928\,\psi_0 > R_*
\simeq 0.577 \,\psi_0.
\end{equation}
Notice that for the largest mass $M$ the modes with wavelength
bigger than $\lambda_{wh}$ are unstable and these modes are larger
than the horizon radius $R_*=\psi_0/\sqrt{3}$. However, because
$R_*$ is not a true causal horizon [i.e., because $f(R \lessgtr
R_*) <0$ has the same signature at both sizes of this horizon
$R_*$].

From the gravitational point of view we see that for physical
wavelengths: $\lambda
>\lambda_{wh}$ the universe is repulsive and suffers an
accelerated expansion, but for $\lambda< \lambda_{wh}$ it is
collapsing. Physical distances with wavelengths in the range
$\lambda_{wh} <\lambda < \psi_0$ are commonly known as
cosmological scales and distances with wavelengths in the range
$0<\lambda < \lambda_{wh}$ are known as astrophysical scales.
These scales we shall refer as the inner of the FWH.

In this work we shall study the cosmological consequences of some
possible big bang produced by a black-hole with mass $M$ in a 5D
extended SdS metric (\ref{a3}). Under these particular
circumstances, the effective 4D metric obtained by the use of a
foliation $\psi=\psi_0=1/H$ is comported as a false white-hole,
which is repulsing all matter which is with $\lambda >
\lambda_{wh}$ and attract modes with $\lambda < \lambda_{wh}$. The
paper is organized as follows: in Sect. II we develop the 5D
dynamics of the scalar field fluctuations of the metric and the
inflaton field in the weak-field approximation. In Sect. III we
explore the effective 4D dynamics of these fields, by using a
semiclassical approximation for the inflaton field. Furthermore,
we calculate the squared field fluctuations for the quantum
inflaton fluctuations and metric fluctuations. These spectrums are
valid outside the FWH. Finally, in Sect. IV we give some final
comments.
\\

\section{5D weak-field limit of the inflation and metric fluctuations}

In order to consider a 5D vacuum on the 5D Ricci flat metric
(\ref{a3}), we shall consider a non-massive scalar field
$\varphi(x^{a})$ which is free of interactions. Its dynamics can
be derived from the action\cite{etgr}
\begin{equation}\label{a8}
^{(5)}{\cal S}=\int \sqrt{g_5}\left[\frac{^{(5)}R}{16 \pi
G}-\frac{1}{2}g^{ab}\varphi_{,a}\varphi_{,b}\right] d^{4}x\,
d\psi,
\end{equation}
where $^{(5)}{\cal R}$ is the Ricci scalar, $g_5$ is the
determinant of the metric (\ref{a3}) and $G$ is the gravitational
constant. The energy-momentum tensor:
$^{(5)}T_{ab}=2\frac{\delta\mathcal{L}}{\delta
g^{ab}}-g_{ab}\mathcal{L}$, derived from the action (\ref{a8}),
reads
\begin{equation}\label{a9}
^{(5)}T_{ab}=\varphi_{,a}\varphi_{,b}-\frac{1}{2}g_{ab}\,\varphi_{,c}\varphi^{,c},
\end{equation}
which is obviously symmetric. The dynamics of the scalar field
$\varphi$ derived from the action (\ref{a8}).

We consider the scalar metric fluctuations
$\Phi(\tau,r,\theta,\phi,\psi)$. In cartesian coordinates this
perturbed line element in the weak-field limit, is
\begin{equation}\label{a7}
\left.dS_{5}^{2}\right|_{pert}=\left(\frac{\psi}{\psi_0}\right)^2
\left[F(\tau,x,y,z) \left[1+{2\Phi}\right]d\tau^2 -
J(\tau,x,y,z)\left[1-{2\Phi}\right]\delta_{ij}
dx^{i}dx^{j}\right]-d\psi^2,
\end{equation}
being now $\Phi=\Phi(\tau,x,y,z,\psi)$. On sufficiently large
scales the following condition is satisfied:
\begin{equation}\label{a17}
\frac{\psi_0}{6\sqrt{3} a(\tau)r} \ll 1,
\end{equation}
where $r_{H}$ denotes the value of the radial coordinate at the
horizon entry and the conformal time $\tau$ is related to the
scale factor by the expression $a(\tau)=-\psi_0/\tau$, so that the
constant Hubble parameter satisfies
\begin{equation}\label{a5}
H=\frac{1}{\psi_0}=a^{-2}\frac{da}{d\tau} .
\end{equation}
On the other hand, one can define the function ${\cal H}=
\dot{a}/a=-1/\tau$. On very large scales, when the weak-field
limit holds, the functions $F(\tau,r)$ and $J(\tau,r)$ become
independent of spatial coordinates [see eqs. in (\ref{a4})], so
that
\begin{equation}
\left.F(\tau,r)\right|_{\frac{\psi_0}{6\sqrt{3} a(\tau)r}\ll 1}
\rightarrow a^2(\tau), \qquad
\left.J(\tau,r)\right|_{\frac{\psi_0}{6\sqrt{3} a(\tau)r} \ll 1}
\rightarrow a^2(\tau),
\end{equation}
and the metric (\ref{a3}) describes an universe which is nearly 3D
spatially homogeneous and  isotropic. In this limit case, the
equation of motion for $\varphi$ can be linearized with respect to
$\Phi$
\begin{equation}\label{a27}
\ddot{\varphi}-(2{\cal
H}-4\dot{\Phi})\dot{\varphi}-(1+4\Phi)\nabla^{2}\varphi+\left(\frac{\psi}{\psi_0}\right)^2a^2
\left[\left(2\overset{\star}{\Phi}-\frac{4}{\psi}\right)\overset{\star}{\varphi}-(1+2\Phi)\overset{\star\star}{\varphi}-\frac{8}{\psi}
\Phi\overset{\star}{\varphi}\right]=0.
\end{equation}
Since now we are dealing with lineal equations of motion for
$\Phi$ and $\varphi$ one can use a semi-classical approximation
for $\varphi$: $\varphi(\tau,x,y,z,\psi)=\varphi_{b}(\tau,\psi)
+\delta\varphi(\tau,x,y,z,\psi)$. From the equation (\ref{a27}),
we obtain separately the dynamics for both, the background part of
the field $\varphi_b$ and the quantum fluctuations $\delta\varphi$
\begin{eqnarray}
\ddot{\varphi}_b &+ & 2{\cal
H}\dot{\varphi}_b-\left(\frac{\psi}{\psi_0}\right)^2a^2\left[\frac{4}{\psi}\overset{\star}{\varphi}_b
+\overset{\star\star}{\varphi}_b\right]=0,\label{a28}\\
\ddot{\delta\varphi}& + & 2{\cal
H}\dot{\delta\varphi}-\nabla^{2}\delta\varphi-\left(\frac{\psi}{\psi_0}\right)^2a^2
\left[\frac{4}{\psi}\overset{\star}{\delta\varphi}+\overset{\star\star}{\delta\varphi}\right]\nonumber
\\
&- &4\dot{\varphi}_{b}\dot{\Phi}
+\left(\frac{\psi}{\psi_0}\right)^{2}a^2\left[2\overset{\star}{\varphi}_{b}\overset{\star}{\Phi}-\left(\frac{8}{\psi}\overset{\star}{\varphi}_{b}
+2\overset{\star\star}{\varphi}_b\right)\Phi\right]=0. \label{a29}
\end{eqnarray}
On the other hand, the background (diagonal) Einstein equations
are
\begin{eqnarray}
3{\cal
H}^2-\frac{3a^2}{\psi_0^2}=\frac{\kappa_5}{2}\left[\dot{\varphi}_{b}^2+\left(\frac{\psi}{\psi_0}\right)^2
a^2 \overset{\star}{\varphi}_b^2
\right], \label{a30}\\
-{\cal H}^{2}-2\dot{{\cal H}}+\frac{3a^{2}}{\psi_{0}^{2}} =
\frac{\kappa_5}{2}\left[\dot{\varphi}_b^2-\left(\frac{\psi}{\psi_0}
\right)^2a^2\overset{\star}{\varphi}_b^2  \right], \label{a31}\\
-3({\cal H}^2+\dot{\cal
H})+\frac{6a^2}{\psi_0^2}=\frac{\kappa_5}{2}\left[\dot{\varphi}_b^2+\left(\frac{\psi}{\psi_0}\right)^2a^2
\overset{\star}{\varphi}_b^2\right], \label{a32}
\end{eqnarray}
while that the linearized fluctuated diagonal Einstein equations
are
\begin{eqnarray}
-6{\cal H}\dot{\Phi}& + &
2\nabla^2\Phi+3a^2\left[\left(\frac{\psi}{\psi_0}\right)^2\overset{\star\star}{\Phi}+\frac{4\psi}{\psi_0^2}\overset{\star}{\Phi}
-\frac{2}{\psi_0^2}\Phi\right] \nonumber \\
& = &
\kappa_5\left[\dot{\varphi}_b\delta\dot{\varphi}+\left(\frac{\psi}{\psi_0}\right)^2a^2
\left(\overset{\star}{\varphi}_b\overset{\star}{\delta\varphi}+\overset{\star}{\varphi}_b^2\Phi\right)\right], \label{a33}\\
2 \ddot{\Phi} & + &6{\cal
H}\dot{\Phi}+a^2\left[\frac{6}{\psi_0^2}\Phi-\left(\frac{\psi}{\psi_0}\right)^2\overset{\star\star}{\Phi}
-\frac{4\psi}{\psi_0^2}\overset{\star}{\Phi}\right] \nonumber \\
&=&
\kappa_5\left[\dot{\varphi}_b\dot{\delta\varphi}-\left(\frac{\psi}{\psi_0}\right)^2a^2
\left(\overset{\star}{\varphi}_b\overset{\star}{\delta\varphi}+\overset{\star}{\varphi}_b^2\Phi\right)\right], \label{a34}\\
3 (\ddot{\Phi} &+ & 4{\cal
H}\dot{\Phi})-\nabla^2\Phi+\frac{6a^2}{\psi_0^2}\left(2\Phi-\psi\overset{\star}{\Phi}\right)
\nonumber \\
& = &\kappa_5
\left[\dot{\varphi}_b\dot{\delta\varphi}+\left(\frac{\psi}{\psi_0}\right)^2a^2\left(\overset{\star}{\varphi}_b\overset{\star}{\delta\varphi}
+\overset{\star}{\varphi}_b^2\Phi\right) \right]. \label{a35}
\end{eqnarray}
On the other hand, linearizing the non-diagonal field Einstein
equations, we obtain
\begin{eqnarray}
2\left({\cal H}\Phi_{,i}+\dot{\Phi}_{,i}\right)&=&\kappa_5\dot{\varphi}_b\delta\varphi_{,i},\label{36}\\
6{\cal
H}\overset{\star}{\Phi}+3\overset{\cdot\star}{\Phi}&=&\kappa_5
\left(\dot{\varphi}_b\overset{\star}{\delta\varphi}+\dot{\delta\varphi}\overset{\star}{\varphi}_b\right), \label{a37}\\
\overset{\star}{\Phi}_{,i}&=&\kappa_5\overset{\star}{\varphi}_b\delta\varphi_{,i}.\label{a38}
\end{eqnarray}
The dynamics of the field $\Phi$ can be described in terms of the
scalar field fluctuations $\delta\varphi$ using a linear
combination of the equations (\ref{a33})-(\ref{a35})
\begin{eqnarray}
\ddot{\Phi}&+&\frac{9}{2}{\cal
H}\dot{\Phi}-\frac{1}{2}\nabla^2\Phi+a^2\left[\frac{9}{2\psi_0^2}\Phi+\frac{\psi}{\psi_0^2}\overset{\star}{\Phi}
+\frac{7}{4}\left(\frac{\psi}{\psi_0}\right)^2\overset{\star\star}{\Phi}\right]
\nonumber \\
&=& \frac{\kappa_5}{4} \left[\dot{\varphi}_b\dot{\delta\varphi}
+9\left(\frac{\psi}{\psi_0}\right)^2
a^2\left(\overset{\star}{\varphi}_b\overset{\star}{\delta\varphi}+\overset{\star}{\varphi}_b^2\Phi\right)\right].
\label{a40}
\end{eqnarray}
Notice that in general the quantum fluctuations $\delta\varphi$
act as a source of scalar metric fluctuations $\Phi$.

\section{Induced 4D dynamics outside the FWH in the weak field limit}

In order to obtain the dynamics for both, $\Phi$ and $\varphi$ on
the effective 4D universe, we shall assume that the 5D spacetime
can be foliated by a family of hypersurfaces
$\Sigma:\psi=constant$. Our 4D universe will be here represented
by a generic hypersurface $\Sigma_{0}:\psi=\psi_0$. Thus, on every
leaf member of the family, the line element induced by (\ref{a3})
has the form
\begin{equation}\label{b1}
ds_4^2=F(\tau,r)d\tau^2-J(\tau,r)[dr^2+r^2(d\theta^2+sin^2\theta
\, d\phi^2)],
\end{equation}
with $F(\tau,r)$ and $J(\tau,r)$ given by (\ref{a4}).

In the weak-field limit, the 5D perturbed line element in
cartesian coordinates (\ref{a7}), induces on the hypersurface
$\Sigma_0$ the effective 4D line element
\begin{equation}\label{b4}
\left.ds_{4}^2\right|_{pert}=F(\tau,\bar{x})\left[1+{2\Omega(\tau,\bar{x})}\right]d\tau^2-J(\tau,\bar{x})\left[1-{2\Omega(\tau,\bar{x})}\right]
\delta_{ij}dx^{i}dx^{j},
\end{equation}
where $\Omega(\tau,\bar{x})\equiv\Phi(\tau,x,y,z,\psi_0)$
describes the 4D scalar metric fluctuations induced on $\Sigma_0$.

The 5D action (\ref{a8}) induces on our 4D spacetime the effective
action
\begin{equation}\label{b10}
^{(4)}{\cal S}_{eff}=\int
d^{4}x\sqrt{g_4}\left[\frac{^{(4)}R}{16\pi
G}-\frac{1}{2}g^{\mu\nu}\bar{\varphi}_{,\mu}\bar{\varphi}_{,\nu}
+V(\bar{\varphi})\right],
\end{equation}
where $g_4$ is the determinant of the 4D induced metric, which for
the background reads $\bar{g}_{4}=-FJ^{3}$ while for the perturbed
metric $g_{4}^{(p)}=-F\,J^3 (1+2\Omega)(1-2\Omega)^3$. In the
linear approximation, the 4D Ricci scalar curvature $^{(4)}R$ is
given by
\begin{equation}\label{b11}
^{(4)}R=\frac{2}{a^2}\left[3 \left({\cal H}^2+\dot{\cal
H}\right)+\nabla^2\Omega - 6 \Omega \left({\cal H}^2+\dot{\cal
H}\right)-3 \left(4{\cal H}\dot{\Omega}+\ddot{\Omega}\right)
\right],
\end{equation}
and the induced 4D effective potential $V$ has the form
\begin{equation}\label{b12}
V(\bar{\varphi})=-\frac{1}{2}g^{\psi\psi}\left.\left(\frac{\partial\varphi}{\partial\psi}\right)^2\right|_{\psi=\psi_0}.
\end{equation}
Thus, let us use the semiclassical approximation:
$\bar{\varphi}(\tau,\vec{x})
=\bar{\varphi}_{b}(\tau)+\delta\bar{\varphi}(\tau,\vec{x})$, where
$\bar{\varphi}_b(\tau)\equiv
\left.\bar{\varphi}_b(\tau,\psi)\right|_{\psi=\psi_0}$ is the
background 4D inflaton field and
$\delta\bar{\varphi}(\tau,\vec{x})\equiv
\left.\delta\bar{\varphi}(\tau,\vec{x},\psi)\right|_{\psi=\psi_0}$
stands for the 4D inflaton field quantum fluctuations. In our
analysis the fields $\Omega$ and $\bar{\varphi}$ are
semi-classical fields, so they are constituted by a classical part
plus a quantum part. To study the dynamics of the former, a
standard quantization procedure will be implemented. To do it, we
shall impose the commutation relations
\begin{equation}\label{b13}
\left[\bar{\varphi}(\tau,\vec{x}),\Pi_{(\bar{\varphi})}^{0}(\tau,\vec{x}^{\prime})\right]=i\delta^{(3)}(\vec{x}
-\vec{x}^{\prime}),\quad
\left[\Omega(\tau,\vec{x}),\Pi_{(\Omega)}^{0}(\tau,\vec{x}^{\prime})\right]=i\delta^{(3)}(\vec{x}-\vec{x}^{\prime}),
\end{equation}
where $\vec{x}$ is denoting the 3D vector position in cartesian
coordinates. Due to the fact that the conjugate momentum to
$\bar{\varphi}$ and $\Omega$ [calculated on the background metric
with determinant $\bar{g}_4=-F\,J^3$],are respectively given by
$\Pi_{(\bar{\varphi})}^{0}=\sqrt{-\bar{g}_4}F^{-1}\dot{\bar{\varphi}}$
and $\Pi_{(\Omega)}^{0}=[12/(16\pi G)]a^{-2}(3\dot{\Omega}-2{\cal
H})\sqrt{-\bar{g}_4}$, the expressions (\ref{b13}) yield
\begin{equation}\label{b14}
\left[\bar{\varphi}(\tau,\vec{x}),\dot{\bar{\varphi}}(\tau,\vec{x}^{\prime})\right]=\frac{i\,F}{\sqrt{-\bar{g}_4}}
\delta^{(3)}(\vec{x}-\vec{x}^{\prime}),\quad
\left[\Omega(\tau,\vec{x}),\dot{\Omega}(\tau,\vec{x}^{\prime})\right]
=i\frac{4\pi G
a^2}{9\sqrt{-\bar{g}_4}}\delta^{(3)}(\vec{x}-\vec{x}^{\prime}).
\end{equation}

\subsection{4D classical dynamics of the inflaton field}

Taking this into account, the evaluation of equation (\ref{a28})
on $\Sigma_0$ yields
\begin{equation}\label{b15}
\ddot{\bar{\varphi}}_b+2{\cal
H}\dot{\bar{\varphi}}_b+a^2m^2\bar{\varphi}_b=0,
\end{equation}
where we have used the relation:
$[(4/\psi)\overset{\star}{\varphi}_b+\overset{\star\star}{\varphi}_b]|_{\psi=\psi_0}
=-m^2\bar{\varphi}_b$, such that $m$ is a separation constant. The
background field $\bar{\varphi}$ must obey the Friedmann-like
equation
\begin{equation}\label{b16}
\left(\frac{\partial\bar{\varphi}_b}{\partial\tau}\right)^2
+a^2\left(\frac{\partial\varphi_b}{\partial\psi}\right)^2_{\psi=\psi_0}=0.
\end{equation}
A particular solution of (\ref{b15}), which also is satisfied when
inflation begins, are the slow rolling conditions:
$\partial\bar{\varphi}_b/\partial\tau=0$, where necessarily $m=0$.
Using this solution in (\ref{b16}), it yields
$\bar{\varphi}_{b}=0$. It means that all the energy density on the
4D hypersurface is induced geometrically by the foliation
$\psi=\psi_0=1/H$, because the background energy density related
to the background inflaton field is null. This is an important
difference with respect to de Sitter models in standard 4D
inflationary models\cite{5} in
which the background energy density is given by the potential. In our case, as can be seen from Eqs. (\ref{a30},\ref{a31}), \ref{a32}),
the right side of these eqs. are zero. Hence, the pressure and energy density being geometrically induced by the foliation:
\begin{eqnarray}
{\cal
H}^2=\frac{a^2}{\psi_0^2}, \label{aa30}\\
{\cal H}^{2}+2\dot{{\cal H}}=\frac{3a^{2}}{\psi_{0}^{2}}, \label{aa31}\\
({\cal H}^2+\dot{\cal H})=\frac{2a^2}{\psi_0^2}, \label{aa32}
\end{eqnarray}
from which we obtain
\begin{equation}\label{ua}
\left( \frac{{\cal H}}{a}\right)^2=\frac{1}{\psi^2_0} =
\frac{\dot{\cal H}}{a}.
\end{equation}
Furthermore, the background induced 4D scalar curvature
$^{(4)}\bar{\cal R}$ is
\begin{equation}\label{ub}
^{(4)}\bar{\cal R} =\frac{2}{a^2}\left[3 \left({\cal
H}^2+\dot{\cal H}\right)\right],
\end{equation}
Using the eq. (\ref{ua}) in (\ref{ub}), we obtain the background
induced scalar curvature
\begin{equation}
^{(4)}\bar{\cal R} = \frac{12}{\psi^2_0},
\end{equation}
which is the exact expression that can be obtained in
STM\cite{wepon} theory for a de Sitter expansion from the
expression
\begin{equation}
^{(4)}\bar{\cal R}= -\frac{1}{4} \left[ \bar{g}^{\mu\nu}_{,4}
\bar{g}_{\mu\nu,4} + \left( \bar{g}^{\mu\nu}
\bar{g}_{\mu\nu,4}\right)^2\right],
\end{equation}
where $\bar{g}_{\mu\nu}$ denotes the background components of the
tensor metric.

\subsection{4D scalar metric fluctuations spectrum}

After working with the equations (\ref{a37}), (\ref{a38}) and
(\ref{a40}), we find that the scalar metric fluctuations $\Omega$
on $\Sigma_0$ obey the equation
\begin{equation}\label{b18}
\ddot{\Omega}+\frac{9}{2}{\cal
H}\dot{\Omega}-\frac{1}{2}\nabla^2\Omega+\left[\frac{9}{2}a ^2
H^2+\lambda^2\right]\Omega=0.
\end{equation}
Here, we have used
$\lbrace(\psi/\psi_0^2-9/\psi)\overset{\star}{\Phi}+[(7/4)(\psi/\psi_0)^2-9/4]\overset{\star\star}{\Phi}\rbrace_{\psi=\psi_0}=\lambda^2\Omega$,
where $\lambda$ is a separation constant with mass units. Now it
can be shown that equation (\ref{36}), evaluated on $\Sigma_0$,
leads to the condition $\dot{\Omega}=-{\cal H}\Omega$. Using this
last condition in equation (\ref{b18}), we obtain
\begin{equation}\label{b19}
\ddot{\Omega}+2{\cal
H}\dot{\Omega}-\frac{1}{2}\nabla^2\Omega+\left[a^2H^2+\lambda^2\right]\Omega=0.
\end{equation}
If we introduce the auxiliary field $\chi(\tau,\bar{r})$, through
the formula $ \Omega(\tau,\bar{r})=e^{-\int{\cal
H}(\tau)d\tau}\chi(\tau,\bar{r})$, the equation (\ref{b19}),
becomes
\begin{equation}\label{b21}
\ddot{\chi}-\frac{1}{2}\nabla^2\chi+\left(\lambda^2-\dot{\cal
H}\right)\chi=0.
\end{equation}
The field $\chi$, can be expanded in terms of the Fourier modes
\begin{equation}\label{b22}
\chi(\tau,\bar{r})=\frac{1}{(2\pi)^{3/2}}\int
d^3k\left[a_ke^{i\bar{k}\cdot\bar{r}}\xi_k(\tau)+a^{\dagger}_{k}\xi_{k}^{*}(\tau)\right],
\end{equation}
where the annihilation and creation operators $a_k$ and
$a_{k}^{\dagger}$, satisfy
\begin{equation}\label{b23}
[a_k,a_{k^{\prime}}^{\dagger}]=\delta^{(3)}(\bar{k}-\bar{k}^{\prime}),\qquad
[a_k,a_{k^{\prime}}]=[a_{k}^{\dagger},a_{k^{\prime}}^{\dagger}]=0.
\end{equation}
Inserting (\ref{b22}) in (\ref{b21}) and using (\ref{b23}), we
find
\begin{equation}\label{b24}
\ddot{\xi}_k+\left(k_{eff}^2-\frac{2}{\tau^2}+\lambda^2\right)\xi_k=0,
\end{equation}
where $k_{eff}^2=k^2/2$. If we require that the modes $\xi_k$ to
be normalized, they must satisfy the following expression on the
UV-sector:
\begin{equation}\label{b25}
\xi_k\dot{\xi}_k^{*}-\xi_k^{*}\dot{\xi}_k=i\frac{4\pi G}{9a_0^2}.
\end{equation}
Thus, choosing the Bunch-Davies vacuum condition, the normalized
solution of (\ref{b24}) reads
\begin{equation}\label{b26}
\xi_k(\tau)\simeq\frac{i\pi}{3a_0}\sqrt{G}{\cal
H}_{\nu}^{(2)}[z(\tau)],
\end{equation}
where ${\cal H}_{\nu}^{(2)}[z(\tau)]$ is the second kind Hankel
function, $\nu=(1/2)\sqrt{1+4\beta}$ and $z(\tau)=k_{eff} \tau$.
Notice that $\beta=2-\lambda^2\tau^2>0$, such that the conformal
time at the beginning of inflation $\tau_0= -{\sqrt{2}\over
\lambda}$ is defined such that $\beta_0=2-\lambda^2\tau^2_0=0$ and
the conformal time at the end of inflation complies with
$\beta_e=2-\lambda^2\tau^2_e \simeq 2$. Using the definition of
$\beta$, it is easy to show that the parameter $\nu\simeq 3/2$ for
$\tau^2_e \ll 2/\lambda^2$, which assures that the spectral index
$1>n_s > 0.96$\cite{rpp}. Using the fact that $1-n_s=3-2\nu$, we
obtain for $\lambda=10^{-10}\, G^{-1/2}$, that the range of
acceptable values for $\tau_e$ is
\begin{equation}
0 < (-\tau_e) < 0.245 \times 10^{10}\, G^{1/2}. \label{ta}
\end{equation}
It is important to see how the universe becomes scale invariant
with the expansion of the universe because $\beta(\tau)$ evolves
from $\beta_0 =0$ to $\beta_e \simeq 2$ along the inflationary
expansion. The temporal evolution of $\beta$, and hence of the
spectral index $n_s$, is due to the existence of $\lambda$, which
has a clear origin in the extra space-like coordinate $\psi$ [see
eqs. (\ref{b18}) and below]. This result cannot be found in
standard 4D inflationary models.\\

The amplitude of the 4D gauge-invariant metric fluctuations $\left<\Omega^2\right>_{IR}$ on the IR-sector ($k_{eff}\tau \ll 1$), is given by
\begin{equation}\label{esp1}
\left<\Omega^2\right>_{IR}=\frac{1}{2\pi^2}\left(\frac{a_0}{a}\right)^2\int_{k=0}^{\epsilon_1 k_{wh}}dk k^2\left. \left(\xi_k\xi_k^{*}\right)\right|_{IR}
\end{equation}
where $\epsilon_1=k_{max}^{IR}/k_p\ll 1$ is a dimensionless parameter, being $k_{max}^{IR}=k_{wh}(\tau_i)=\sqrt{(4/\tau_i^2)-2\lambda^2}$
the wave number related to the Hubble radius by the time when the modes re-enter to the horizon $\tau_i$. The Planckian wave number is
here denoted by $k_p$. Now, considering the asymptotic expansion for the Hankel function ${\cal H}_{\nu}^{(2)}[x]
\simeq -(i/\pi)\Gamma(\nu)(x/2)^{-\nu}$ in the expression (\ref{b26}), the equation (\ref{esp1}) yields
\begin{equation}\label{esp2}
\left<\Omega^2\right>_{IR}=\frac{2^{1+3\nu}}{3^2}\Gamma^{2}(\nu)\left(\frac{H
\tau}{2\pi}\right)^2\left(\frac{1}{aH}\right)^{2-2\nu}\int_{0}^{\epsilon_1
k_{wh}} \frac{dk}{k} k^{3-2\nu}.
\end{equation}
It can be easily seen from this equation that the corresponding spectrum for scalar metric fluctuations reads
\begin{equation}\label{esp3}
{\cal
P}_{\Omega}(k)=\frac{2^{1+3\nu}}{3^2}\Gamma^{2}(\nu)\left(\frac{H\tau}{2\pi}
\right)^2\left(\frac{1}{aH}\right)^{2-2\nu}k^{3-2\nu}.
\end{equation}
This spectrum results nearly scale invariant for $\nu\simeq 3/2$, value that may be achieved when $\lambda^2\tau^2\ll 1$.
Performing the integration in (\ref{esp2}), the amplitude of the scalar metric fluctuations is given finally by
\begin{equation}\label{esp4}
\left<\Omega^2\right>_{IR}=\frac{2^{1+3\nu}}{3^2}\Gamma^{2}(\nu)\left(\frac{H\tau}{2\pi}
\right)^2\left(\frac{1}{aH}\right)^{2-2\nu}\epsilon_{1}^{3-2\nu}k_{wh}^{3-2\nu},
\end{equation}
which tends to zero as $\tau \rightarrow 0$.

\subsection{4D inflaton field fluctuations}

Since the background inflaton field $\varphi_b$, is a constant,
the dynamics of the inflaton field fluctuations $\delta\varphi$
are given by the first row of the equation of motion (\ref{a29})
on the hypersurface $\psi=\psi_0$. The equation of motion for the
time dependent modes $\zeta_k(\tau)$ is
\begin{equation}\label{z5}
\ddot{\zeta}_k+ 2 {\cal H} \,\dot\zeta_k+ \left[ k^2 - a^2
\lambda^2 \right] \zeta_k(\tau)=0,
\end{equation}
which has the general solution
\begin{equation}\label{zz5}
\zeta_{k}(\tau)= A_1 \, \left(-\tau\right)^{3/2} {\cal
H}^{(1)}_{\mu}\left[-k\,\tau\right] + A_2 \,
\left(-\tau\right)^{3/2} {\cal
H}^{(2)}_{\mu}\left[-k\,\tau\right],
\end{equation}
Here, ${\cal H}^{(1,2)}_{\mu}[-k\tau]$ are respectively the first
and second kind Hankel functions, $\mu=
\sqrt{9+4\lambda^2\psi^2_0}/2$. One can define $\zeta_k(\tau) =
\tau\, \sigma_{k}(\tau)$, such that $\sigma_k(\tau)$ are the time
dependent modes for the redefined field fluctuations
$\sigma(\vec{x},\tau)$, that can be expanded in terms of Fourier
modes
\begin{equation}\label{b222}
\sigma(\tau,\vec{x})=\frac{1}{(2\pi)^{3/2}}\int
d^3k\left[A_ke^{i\vec{k}\cdot\vec{x}}\sigma_k(\tau)+A^{\dagger}_{k}e^{-i\vec{k}\cdot\vec{x}}\sigma_{k}^{*}(\tau)\right],
\end{equation}
where the annihilation and creation operators $A_k$ and
$A_{k}^{\dagger}$ satisfy the commutation algebra
\begin{equation}\label{b223}
[A_k,A_{k^{\prime}}^{\dagger}]=\delta^{(3)}(\vec{k}-\vec{k}^{\prime}),\qquad
[A_k,A_{k^{\prime}}]=[A_{k}^{\dagger},A_{k^{\prime}}^{\dagger}]=0,
\end{equation}
and to be fulfilled the algebra
\begin{equation}
\left[\sigma(\tau,\vec{x}),\dot\sigma(\tau,\vec{x}')\right]=i\,\delta^{(3)}\left(\vec{x}-\vec{x}'\right),
\end{equation}
must require the normalization condition $\sigma_k \dot\sigma_k^*
- \dot\sigma_k \sigma^* = i$. Therefore, the normalization
constants are given by
\begin{equation}\label{zz51}
A_2 = -\frac{\sqrt{\pi}}{2\psi_0}\,e^{-i \nu \pi/2}, \qquad A_1 =
0.
\end{equation}
The amplitude of the fluctuations of the inflaton field on large scales $(k\tau\ll 1)$, is obtained by means of the formula
\begin{equation}\label{inf1}
\left<\delta\varphi^2\right>_{IR}=\frac{1}{2\pi^2}\left(\frac{a_0}{a}\right)^2\int_{k=0}^{\epsilon_2 k_{wh}}dk k^2\left.
\left(\sigma_{k}\sigma_{k}^{*}\right)\right|_{IR},
\end{equation}
where $\epsilon_{2}=k_{max}^{IR}/k_p$, being in this case $k_{max}^{IR}=\sqrt{(2/\tau_i)+a_{i}^2\lambda^2}$, with $a_i=a(\tau_i)$.
Thus, making use of the asymptotic expansion for the Hankel function ${\cal H}_{\mu}^{(2)}[x]\simeq -(i/\pi)\Gamma(\mu)(x/2)^{-\mu}$,
the expression (\ref{inf1}) becomes
\begin{equation}\label{inf2}
\left<\delta\varphi\right>_{IR}=\frac{a_0^22^{2\mu-1}}{\psi_{0}^2}\frac{\Gamma^2(\mu)}{\pi}\left(\frac{H}{2\pi}\right)^2
\left(\frac{1}{aH}\right)^{3-2\mu}\int_{0}^{\epsilon_{2}k_{wh}}\frac{dk}{k}k^{3-2\mu}.
\end{equation}
After straightforward calculations, the mean squared fluctuations for the inflaton field read
\begin{equation}\label{inf3}
\left<\delta\varphi^2\right>_{IR}=\frac{a_0^22^{2\mu-1}}{\psi_0^2}\frac{\Gamma^{2}(\mu)}{\pi}\left(\frac{H}{2\pi}\right)^2\epsilon_{2}^{3-2\mu}\left(\frac{k_{wh}}{aH}\right)^{3-2\mu}.
\end{equation}
In this manner, the spectrum for the inflaton fluctiations on 4D, is given by
\begin{equation}\label{inf4}
{\cal P}_{\delta\varphi}(k)=\frac{a_0^22^{2\mu-1}}{\psi_0^2}\frac{\Gamma^{2}(\mu)}{\pi}\left(\frac{H}{2\pi}\right)^2\left(\frac{k}{aH}\right)^{3-2\mu}.
\end{equation}
This spectrum shows that its nearly scale invariance is achieved for $\mu\simeq 3/2$, and in turn this value is obtained when $\lambda^2\psi_{0}^2\ll 1/4$.

\section{Final Comments}

We have studied an interesting case in which the universe expands
on an effective 4D hypersurface obtained after take a constant
foliation on the extra space-like coordinate of an extended SdS
Ricci-flat metric. The 4D effective space-time describes an
universe that expands on cosmological scales but collapse on
astrophysical ones because it has a black-hole in its center with
mass $M=\psi_0/83\sqrt{3} G)$.  Under these circumstances, the
universe behaves as a white hole that evaporates on scales greater
its radius. The behavior of the universe on these large scales is
similar to that a white hole, so that we have called it a false
white hole (FWH). We have studied the large scales evolution of
the scalar metric fluctuations and the quantum fluctuations of the
metric. In both cases the amplitude of the fluctuations decreases
during inflation for power spectrums that are nearly scale
invariant on cosmological scales.

\section*{Acknowledgements}

\noindent J. E. Madriz-Aguilar, and C. Moreno acknowledge CONACYT
(Mexico) and Mathematics Department of CUCEI- UdG for financial
support. M. Bellini acknowledges CONICET (Argentina) and UNMdP for
financial support.

\end{document}